\documentclass{article}
\usepackage[utf8]{inputenc}
\usepackage{graphicx}
\usepackage{amssymb}
\usepackage{amsmath}
\usepackage{amsthm}
\usepackage{amsfonts,amsmath}
\usepackage{tikz}
\usetikzlibrary{shapes,arrows}
\usetikzlibrary{arrows,positioning}
\usepackage[none]{hyphenat} 
\usepackage{cite}
\usepackage{hyperref}
\usepackage{colortbl}
\usepackage{adjustbox}
\usepackage{colortbl}
\usepackage{fancybox}
\usepackage{dcolumn}
\usepackage{multirow}
\usepackage{tabularx}
\usepackage{enumitem}
\usepackage{float}
\usepackage{soul}
\usepackage[pagewise]{lineno}
\usepackage{siunitx}
\usepackage{arydshln}
\usepackage{authblk}

\usepackage{verbatim,caption}
\usepackage{listings}

\usepackage[textsize=tiny]{todonotes}

\usepackage{xcolor,cancel}

\usepackage[pagewise]{lineno}

\newtheorem{Proposition}{Proposition}[section]

\newtheorem{Definition}{Definition}[section]

\numberwithin{equation}{section}

\newcommand{\bfa}[1]{\mbox{\boldmath $ #1 $}}
\newcommand{\D}{\displaystyle}

\begin{document}

\title{Structural and Practical Identifiability of Phenomenological Growth Models for Epidemic Forecasting}

\author[1]{Yuganthi R. Liyanage\textsuperscript{*}}
\author[2,3]{Gerardo Chowell\textsuperscript{*}}
\author[4]{Gleb Pogudin}
\author[1]{Necibe Tuncer}

\affil[1]{Department of Mathematics and Statistics, Florida Atlantic University, Boca Raton, Florida, USA}
\affil[2]{School of Public Health, Georgia State University, Atlanta, Georgia, USA}
\affil[3]{Department of Applied Mathematics, Kyung Hee University, Yongin 17104, Korea}
\affil[4]{LIX, CNRS, Ecole polytechnique, Institute Polytechnique de Paris, Paris, France}

\affil[*]{Joint corresponding authors: Yuganthi R. Liyanage (\href{mailto:aliyanage2018@fau.edu}{aliyanage2018@fau.edu}) and Gerardo Chowell (\href{mailto:gchowell@gsu.edu}{gchowell@gsu.edu})}

\maketitle
\abstract{Phenomenological models are highly effective tools for forecasting disease dynamics using real-world data, particularly in scenarios where detailed knowledge of disease mechanisms is limited. However, their reliability depends on the model parameters' structural and practical identifiability. In this study, we systematically analyze the identifiability of six commonly used growth models in epidemiology: the generalized growth model (GGM), the generalized logistic model (GLM), the Richards model, the generalized Richards model (GRM), the Gompertz model, and a modified SEIR model with inhomogeneous mixing. To address challenges posed by non-integer power exponents in these models, we reformulate them by introducing additional state variables. This enables rigorous structural identifiability analysis using the StructuralIdentifiability.jl package in JULIA. We validate the structural identifiability results by performing parameter estimation and forecasting using the \textit{GrowthPredict} MATLAB toolbox. This toolbox is designed to fit and forecast time-series trajectories based on phenomenological growth models. We applied it to three epidemiological datasets: weekly incidence data for monkeypox, COVID-19, and Ebola. Additionally, we assess practical identifiability through Monte Carlo simulations to evaluate parameter estimation robustness under varying levels of observational noise. Our results confirm that all six models are structurally identifiable under the proposed reformulation. Furthermore, practical identifiability analyses demonstrate that parameter estimates remain robust across different noise levels, though sensitivity varies by model and dataset. These findings provide critical insights into the strengths and limitations of phenomenological models to characterize epidemic trajectories, emphasizing their adaptability to real-world challenges and their role in informing public health interventions.}


\section{Introduction}

Phenomenological growth models are widely used to describe infectious disease dynamics, offering critical insights into the parameters governing their spread and control \cite{Bürger2021}.
A critical aspect for ensuring the reliability of epidemiological modeling is the identifiability of the model parameters \cite{Miao2008,tuncer2016RiftValley,Eisenberg2013,Simpson2022,chowell2024investigating}, as it determines whether the values of the parameters can be accurately and uniquely estimated from available data. Structural identifiability analysis addresses whether model parameters can be uniquely determined from perfect, error-free data \cite{tuncer2016RiftValley}, which forms the basis for robust modeling. Without this assurance, reliable parameter estimation becomes impossible, undermining the model's utility for forecasting and informing public health interventions.

Structural identifiability is particularly challenging for phenomenological models, which often involve nonlinear dynamics and transformations with non-integer exponents that complicate traditional analytical approaches. In this study, we focus on analyzing the structural identifiability of six widely adopted growth models in epidemiology: the generalized growth model (GGM), generalized logistic model (GLM), Richards model, generalized Richards model (GRM), Gompertz model, and a modified SEIR model incorporating inhomogeneous mixing \cite{Viboudll2016}. These models capture a range of growth patterns, from sub-exponential to logistic growth, enabling application to diverse epidemiological settings.

We use the differential algebra approach implemented in the StructuralIdentifiability.jl package in JULIA \cite{structidjl2023} to assess structural identifiability. This approach eliminates unobserved state variables and derives differential algebraic polynomials that relate observed variables and model parameters. For models with non-integer exponents, we reformulate them by introducing additional state variables, ensuring compatibility with existing analysis methods. This adaptation enables identifiability analysis for models that would otherwise be analytically intractable \cite{Margaria2004,Margaria0011}.

However, real-world data often contains noise and measurement errors, making some parameters difficult to estimate in practice. This is where practical identifiability becomes important. Practical identifiability analysis evaluates whether structurally identifiable parameters can still be estimated with reasonable accuracy given real-world observational noise \cite{miao2011,Liyanage2024,tuncer2018structural}. Practical identifiability analysis allows us to evaluate the extent to which structurally identifiable parameters can be estimated from noisy observations. By simulating datasets with varying noise levels, we examine each model's robustness and the practical feasibility of accurately estimating its parameters. To evaluate these models' performance and practical identifiability under real-world conditions, we used the \textit{GrowthPredict} MATLAB toolbox \cite{toolbox2024}, which facilitates parameter estimation and forecasting using a suite of phenomenological growth models.

The contributions of this work are twofold: 1) we rigorously establish the structural identifiability of commonly used phenomenological models, providing theoretical assurance for parameter estimation; and 2) we investigate the practical identifiability of these models under noisy data conditions, offering insights into their reliability and constraints when applied to real-world epidemic datasets. These findings build upon and complement prior works that have established identifiability for other epidemiological models, including compartmental and mechanistic frameworks, thereby situating our work within the broader literature on model identifiability \cite{BELLMAN1970,Roosa2019,Sauer2022,MASSONIS2021,Tuncer31122022,Gallo2022,Renardy2022,EVANS2005,Chowell2023}. Together, these contributions enhance the utility of phenomenological models for forecasting infectious disease dynamics and guiding public health decision-making. The input codes and output results from JULIA and Matlab used in this paper are publicly accessible on our GitHub repository: \url{https://github.com/YuganthiLiyanage/Phenomenological-Growth-Models}.

\section{Structural Identifiability of Phenomenological Models}

Structural identifiability analysis is a theoretical framework used to determine whether the parameters of a model can be uniquely inferred from unlimited, error-free observations. This analysis is essential for ensuring the mathematical integrity of a model before applying it to real-world data, as it confirms that parameter estimates are both unique and reliable. Various techniques, such as differential algebra methods, the Taylor series expansion method, and input-output approaches, have been developed to evaluate structural identifiability \cite{miao2011,tuncer2018structural}.

In this study, we employ the differential algebra method, a powerful approach that systematically eliminates unobserved state variables and derives differential algebraic polynomials involving the observed variables and model parameters \cite{tuncer2018structural,Chowell2023,Tuncer2018_zika}. To achieve this, we utilize the StructuralIdentifiability.jl package in JULIA, which enables efficient derivation of these polynomials and facilitates rigorous identifiability analysis \cite{structidjl2023}.

We reformulate the model structure for models with non-integer power exponents by introducing additional state variables. This reformulation not only preserves the fundamental structure of the model but also extends the applicability of identifiability analysis to previously intractable cases. By applying this approach, we ensure that parameters and state variables remain identifiable across models~\ref{ModelM1} to~\ref{ModelM6}, broadening the utility of these models in practical epidemic forecasting.

\textbf{Structural identifiability theory}
To analyze the structural identifiability of Models \ref{ModelM1} through \ref{ModelM6}, we rewrite them in the following compact form,
\begin{equation} \label{compact_form}
\begin{aligned}
x'(t)&=f(x, \bfa p), \\ y(t)&=g(x,\bfa p),
\end{aligned}
\end{equation}
where $x(t)$ denotes the vector of state variables, $y(t)$ represents the observations, and $\bfa p$ denotes the vector of parameters. 
A parameter $p$ in a model~\eqref{compact_form} is called structurally identifiable if its value can be uniquely determined from the observable $y(t)$ (assuming noise-free measurements).
A similar property for states is typically referred to as observability but in the context of the present paper it will be convenient for us to use the term identifiability in both cases.
In fact, we will define identifiability for an arbitrary function of parameters and states.
We will give a simplified version of the definition avoid unnecessary technicalities, a formalized version can be found in~\cite[Definition~2.5]{Hong2020}.

\begin{Definition}\label{Def:SI}
A function $h(x, \bfa p)$ in the states and parameters of model~\eqref{compact_form}  is said to be structurally globally identifiable if, for generic $(x, \bfa p)$ and any $(\hat{x}, \hat{\bfa p})$, the following implication holds:
\[
g(x,\bfa p)=g(\hat{x},\hat{\bfa p}) \quad\textit{implies}\quad h(x, \bfa p) = h(\hat{x}, \hat{\bfa p}).
\]
If all the parameters are identifiable, then we will say that the model is identifiable.
\end{Definition}

\begin{Definition}
A function $h(x, \bfa p)$ in the states and parameters of model~\eqref{compact_form}  is said to be structurally locally identifiable if, for generic $(x, \bfa p)$, there exists a neighborhood such that, for any $(\hat{x}, \hat{\bfa p})$, the following implication holds:
\[
g(x,\bfa p)=g(\hat{x},\hat{\bfa p}) \quad\textit{implies}\quad h(x, \bfa p) = h(\hat{x}, \hat{\bfa p}).
\]
\end{Definition}

While the definitions above reflect the intuitive notion of identifiability, when it comes to checking identifiability for specific models, they are not very convenient.
One standard approach to assessing structural identifiability is via input-output equations (also referred to as the differential algebra approach).
We will not describe input-output equations in full generality but only for single-output model since all the models considered in this paper belong to this class.
For a single-output model, the input-output equation is the irreducible equation of minimal order satisfied by the output.
We will normalize the equation by dividing by the leading coefficient (considered as a polynomial in $y$ and its derivatives) to obtain a monic polynomial.
Let $c_1(\bfa p), \ldots, c_\ell(\bfa p)$ be the coefficients of this equation. 
Then, under a certain assumption on the Wronskian of the input-output equations (see, e.g., \cite[Lemma~4.6]{Ovchinnikov2023}), identifiable functions of parameters are precisely the ones expressible in terms of $c_1, \ldots, c_\ell$.
In this paper, we use software \texttt{StructuralIdentifiability.jl} to assess identifiability, and the software automatically checks if the aforementioned condition on Wronskians is fulfilled.

\textbf{Structural identifiability results for the generalized growth model} 

We derive input-output equations by eliminating unobserved state variables using the differential algebra method. This process generates algebraic relationships involving the observed variables and model parameters, enabling identifiability analysis. For example, the generalized growth model is defined by the following differential equation.
\begin{equation}\begin{split}
\textbf{GGM: \hspace{1 cm}}
&\D\frac{dC}{dt}= r C^\alpha(t)
\label{eq:model1}
\end{split}\end{equation}
where, $C'(t)$ denotes incidences at time $t$, $C(t)$ denotes the cumulative number of cases at time $t$ and $\alpha$ denotes the growth rate of the infectious diseases such that $0\leq \alpha \leq1.$ Since epidemic data is typically collected as incident case counts over discrete time intervals, we assume that the incidence, $C'(t)$,corresponds to the observed data. This assumption is made for methodological consistency and computational feasibility, as it simplifies the identifiability analysis by ensuring that our model structure directly aligns with real-world epidemic datasets. It also allows us to use a standardized observation function across different growth models, facilitating direct comparisons in both structural and practical identifiability assessments.

To facilitate this process, we introduce an additional state variable, which allows us to reformulate models with non-integer power exponents into a structure that can be analyzed using the differential algebra approach. For this purpose, we introduce an additional state variable $x(t)$ to eliminate the non-integer power exponent, resulting in the extended version of the model by letting $x(t)= r C^\alpha(t)$, then \textbf{GGM} becomes, 
\begin{equation}\label{ModelM1} \tag{M1}
\begin{aligned}
\D\frac{dC}{dt}&= x(t)\\
\D\frac{dx}{dt}&= \alpha \frac{x^2(t)}{C(t)}.
\end{aligned}
\end{equation}

In this study, we examine the scenario in which the observations are equal to the incidences, that is $y(t)=x(t)$.
The monic input-output equation obtained from \textit{StructuralIdentifiability.jl} package in JULIA, with the observation $y(t)$ is given by,
$$0=-y'' y \alpha + y'^2 (2 \alpha - 1)= y'' y + y'^2 \left(2 - \D\frac{1}{\alpha}\right)$$

In the context of structural identifiability, the input-output equation indicates that $2 - 1/\alpha$
can be identified with the observation $y(t)$, so $\alpha$ is identifiable as well. 
Now, we can express $C(t)= \alpha \frac{x^2(t)}{x'(t)}$  in terms of identifiable parameters and state variables. As a result, $C(t)$ becomes identifiable. This implies that the parameter $r$ is identifiable. Therefore, model \eqref{eq:model1} is structurally identifiable with the observation $y(t).$ We state the following proposition.

\begin{Proposition} \label{prop1}
The generalized growth model \textbf{GGM} is structurally identifiable from the observation of incidences,  $y= x(t)$.
\end{Proposition}

We would like to stress that the way the lifting to the extended model~\eqref{ModelM1} is performed is important to get correct identifiability results for the original model.
Suppose that, instead of $x(t) = rC^\alpha(t)$, we would have introduced $\tilde{x}(t) = C^\alpha(t)$.
Then we would get a different extended system
\begin{equation}\label{eq:extended_too_much}
\D\frac{dC}{dt} = r \tilde{x}(t)\quad\text{ and }\quad \D\frac{d\tilde{x}}{dt} = \alpha r \frac{\tilde{x}^2(t)}{C(t)} 
\end{equation}
In this model, neither $r$ not $\tilde{x}$ are identifiable because there exists an output-preserving transformation $r \to \lambda r, \; \tilde{x} \to \frac{\tilde{x}}{\lambda}$ for any nonzero number $\lambda$.
The discrepancy between the identifiability results for different extended models can be explained as follows.
The trajectories of the original model~\eqref{eq:model1} and of~\eqref{ModelM1} are parametrized by three numbers: $\alpha, r, C(0)$ for the former and $\alpha, C(0), x(0)$ for the latter.
On the other hand, the space of trajectories of~\eqref{eq:extended_too_much} is four-dimensional parametrized by $\alpha, r, C(0), \tilde{x}(0)$.
The images of trajectories of~\eqref{eq:model1} among the trajectories of~\eqref{eq:extended_too_much} are constrained to a manifold $\tilde{x}(t) - C^\alpha(t) = 0$.
And thus this is the additional trajectories which~\eqref{eq:extended_too_much} possesses which turn $r$ into nonidentifiable.

\textbf{Structural identifiability results for the generalized logistic growth model} 

The generalized logistic growth model is given by the following equation.
\begin{equation}\begin{split}
\textbf{GLM: \hspace{1 cm}}
&\D\frac{dC}{dt}= r C^\alpha(t) \left(1-\D\frac{C(t)}{k}\right)
\label{eq:model2}
\end{split}\end{equation}
where $r$ is the generalized growth rate, $k$ is the final epidemic size, $C'(t)$ denotes incidences at time $t$, $C(t)$ denotes the cumulative number of cases at time $t$ and the parameter $\alpha \in [0,1]$ denotes the different growth scenarios; the constant incidents $\alpha=0$, sub-exponential growth $0<\alpha<1$ and exponential growth $\alpha=1$. To obtain the extended version of the model without non integer exponent, we substitute $x(t)=r C^\alpha(t)$, \textbf{GLM} becomes,
\begin{equation}\label{ModelM2}\tag{M2}
\begin{aligned}
\D\frac{dC}{dt}&=  x(t)\left(1-\frac{C(t)}{k}\right)\\
\D\frac{dx}{dt}&=  \alpha  \frac{x^2(t)}{C(t)}\left(1-\frac{C(t)}{k}\right).
\end{aligned}
\end{equation}

Here, we consider the case where the observations $y(t)= x(t) \left(1- \frac{C(t)}{k}\right)$, correspond to the incidences. The
input-output equation obtained from JULIA and normalized, with the observation $y(t)$ is given by,

\begin{equation} \label{Eq_inputoutputM2} 
    \begin{aligned}
0&= y''^2 y^2 + y'' y'^2 y \frac{- 3 \alpha + 1}{\alpha} + 2 y''
y' y^3 \frac{ - \alpha^2 + 1}{k\alpha} + y''  y^5\frac{\alpha^3 + 3 \alpha^2 + 3 \alpha + 1}{\alpha k^2} + y'^4 \frac{2 \alpha -
 1}{\alpha} +\\& 2 y'^3 y^2 \frac{2 \alpha^2 - \alpha - 1}{k\alpha} + y'^2 y^4\frac{- 2 \alpha^3 - 5 \alpha^2 - 4 \alpha - 1}{\alpha k^2}
  \end{aligned}
\end{equation}
We will check that all the parameters can be expressed in terms of the coefficients of this equations, thus, showing that they are identifiable.
First, $\alpha$ can be expressed from $\frac{2\alpha - 1}{\alpha}$, so it is identifiable.
Next, $k$ can be expressed from $\alpha$ and the coefficient $\frac{-\alpha^2 + 1}{\alpha k}$. This concludes that only parameters $k$ and $\alpha$ are identifiable. 

Using \textit{StructuralIdentifiability.jl} package in JULIA software, we obtained the state variable $x(t)$ and $C(t)$ are identifiable (observable) from the given observation. Then $r = \frac{x(t)}{C^{\alpha}(t)}$ can be written as a combination of identifiable parameters and the state variables. Therefore, the parameter $r$ is also identifiable. We assert the following proposition.

\begin{Proposition} \label{prop2}
The generalized logistic growth model~\eqref{eq:model2} is structurally identifiable from the observation of incidences,  $y(t)= x(t)\left(1-\frac{C(t)}{k}\right)$.
\end{Proposition}

\textbf{Structural identifiability results for Richards model} 

The Richards model is given by the following equation. 
\begin{equation}\begin{split}
\textbf{Richards: \hspace{1 cm}}
&\D\frac{dC}{dt}= r C(t) \left(1-\left(\D\frac{C(t)}{k}\right)^a\right)
\label{eq:model3}
\end{split}\end{equation}

We let $x(t)=(\frac{C(t)}{k})^a$, \textbf{Richards} becomes, 
\begin{equation}\label{ModelM3}\tag{M3}
\begin{aligned}
\D\frac{dC}{dt}&= rC(t)\left(1-x(t)\right)\\
\D\frac{dx}{dt}&= a r x(t)\left(1-x(t)\right).
\end{aligned}
\end{equation}

Here, we consider the case when the observations, $y(t)=rC(t)\left(1-x(t)\right)$, correspond to the incidences.
The normalized input-output equation of the model~\eqref{ModelM3} is given by,
\begin{equation*}
\begin{aligned}
 0&=- y''y + y'^2\frac{2a + 1}{a + 1} + y' y \frac{a r(a - 1)}{a + 1} - y^2 \frac{a^2 r^2}{a + 1}
 \end{aligned}
\end{equation*}
The parameter $a$ can be expressed from the coefficient $\frac{2a + 1}{a + 1} = 2 - \frac{1}{a + 1}$.
Next, parameter $r$ can be expressed from $a$ and coefficient $\frac{ar(a - 1)}{a + 1}$.
Thus, both $a$ and $r$ are identifiable from the observation $y(t)$.

Now we check the identifiability of the state variables, obtain that both $x(t)$ and $C(t)$ are identifiable from the observation. Therefore, parameter $a$ can be written as a combination of the identifiable parameters and state variables. Thus, $a$ is identifiable. We state the following proposition.

 \begin{Proposition} \label{prop3}
Richards model~\eqref{eq:model3} is structurally identifiable from the observation of incidences, $y(t)=rC(t)\left(1-x(t)\right)$.
\end{Proposition}

\textbf{Structural identifiability results for the generalized Richards model} 

The model is represented by the following equation

\begin{equation}\begin{split}
\textbf{GRM: \hspace{1 cm}}
&\D\frac{dC}{dt}= r C^p(t) \left(1-\left(\D\frac{C(t)}{k}\right)^a\right)
\label{eq:model4}
\end{split}\end{equation}
where $r$ is the generalized growth rate, $k$ is the final epidemic size, $C'(t)$ denotes incidences at time $t$, $C(t)$ denotes the cumulative number of cases at time $t$ and the parameter $\alpha \in [0,1]$ denotes the different growth scenarios; the constant incidents $\alpha=0$, sub-exponential growth $0<\alpha<1$ and exponential growth $\alpha=1$, and the exponent $a$ denotes the deviation from the symmetric s-shaped dynamics of the simple logistic curve.

To obtain the extended version, without any non-integer power exponent, we let $x(t)= r C^p(t) \quad\textit{and}\quad z(t)= (C/k)^a(t)$, \textbf{GRM} becomes,
\begin{equation}\label{ModelM4}\tag{M4}
\begin{aligned}
\D\frac{dC}{dt}&= x(t)\left(1- z(t)\right)\\
\D\frac{dx}{dt}&= \alpha  \frac{x^2(t)}{C(t)}\left(1- z(t)\right)\\
\D\frac{dz}{dt}&= a  \frac{z(t) x(t)}{C(t)}\left(1- z(t)\right),
\end{aligned}
\end{equation}

Using \textit{StructuralIdentifiability.jl} in JULIA, we determined that the parameters $a$ and $\alpha$ are locally identifiable, and the state variables $x(t)$ and $z(t)$ are also locally identifiable. Furthermore, the product $a^2$ and the summation $a +2 \alpha$ are globally identifiable. Since $a$ is positive, it follows that $a$ is globally identifiable, ensuring that $\alpha, x(t),$ and $z(t)$ are globally identifiable as well.
To obtain the identifiability of parameters $r$ and $k$, we express $r= \D\frac{x(t)}{C^\alpha(t)}$ and $k= \sqrt[a]{\D\frac{C^a(t)}{z(t)}}$ in terms of identifiable states and parameters. Therefore, the parameters $r$ and $k$ are identifiable. We conclude the following proposition.

\begin{Proposition} \label{prop4}
The generalized Richards model \eqref{eq:model4} is structurally identifiable from the observation of incidences, $y(t)=x(t)\left(1-z(t)\right)$ under the assumption of positivity of $a$ and $k$.
\end{Proposition}

\textbf{Structural identifiability results for Gompertz model} 

The Gompertz model is given by the following equation.
\begin{equation}\begin{split}
\textbf{GOM: \hspace{1 cm}}
&\D\frac{dC}{dt}= r C(t) e^{-bt} 
\label{eq:model5}
\end{split}\end{equation}

Letting, $x(t)= r e^{-bt}$, \textbf{GOM} becomes,
\begin{equation}\label{ModelM5}\tag{M5}
\begin{aligned}
\D\frac{dC}{dt}&= C(t) x(t)\\
\D\frac{dx}{dt}&= -be^{-bt}=-b x(t).
\end{aligned}
\end{equation}

Here, we consider the case when the
observations, $y(t)$, correspond to the incidences, $y(t)=C(t)x(t).$

The input-output equation of model \eqref{ModelM5} is given by,
\begin{equation*}
    \begin{aligned}
0&=- y'' y + y'^2 - y' y b - y^2 b^2
    \end{aligned}
\end{equation*}
Since $b$ appears among the coefficients, it is identifiable.
\texttt{Structural Identifiability} package in JULIA shows that both $C(t)$ and $x(t)$ are identifiable from the observation $y(t)$. Thus, the parameter $r= x(t) e^{bt}$ can be written as a product of identifiable parameters and state variables. Thus, $r$ is identifiable.

\begin{Proposition} \label{prop5}
The model~\eqref{eq:model5} is structurally identifiable from the observation of incidences, $y(t)= C(t)x(t)$.
\end{Proposition}

\textbf{Structural identifiability results for SEIR model with inhomogeneous mixing} 

SEIR model with inhomogeneous mixing is given by the following system of equations,

\begin{equation}\begin{split}
\textbf{SEIR \hspace{1 cm}}
&\D\frac{dS}{dt}= \frac{-\beta S(t) I(t)^\alpha}{N} \\
&\D\frac{dE}{dt}= \frac{\beta S(t) I(t)^\alpha}{N} -k E(t) \\
&\D\frac{dI}{dt}= k E(t) -\gamma I(t)\\
&\D\frac{dR}{dt}= \gamma I(t).
\label{eq:model6}
\end{split}\end{equation}

Let $x(t)= \beta I^\alpha(t)$ then \textbf{SEIR} becoms,

\begin{equation}\label{ModelM6}\tag{M6}
\begin{aligned}
\D\frac{dS}{dt}&= \frac{- S(t) x(t)}{N}\\
\D\frac{dE}{dt}&= \frac{S(t) x(t)}{N} -k E(t) \\
\D\frac{dx}{dt}&= \frac{\alpha x(t)}{I(t)} (k E(t) -\gamma I(t))\\
\D\frac{dI}{dt}&= k E(t)-\gamma I(t)\\
\D\frac{dR}{dt}&= \gamma I(t).
\end{aligned}
\end{equation}

Here, we consider the case when the observations are $y(t)= k E(t)$. Using \textit{Structural Identifiability} in JULIA, we determine that the parameters $\alpha$, $\gamma$, and $k$ are globally identifiable, while $N$ is not identifiable. However, since $N$ represents the total population size of a closed population, its value is known, and $N = S_0+E_0+I_0+R_0$. 
With $N$ being a known quantity, we treat it as an additional observation in the model. 
We then perform the identifiability analysis again and obtain the state variable $x(t)$ is identifiable from the observations $y(t)$ and $N$.
Thus, we can express $\beta= \frac{x(t)}{I^\alpha}$ as a combination of identifiable parameters and state variables.
Thus, model \eqref{eq:model6} is structurally identifiable. 

In cases with time-varying $N(t)$, it introduces additional degrees of freedom into the model, potentially leading to unidentifiability of key parameters unless additional constraints or external data sources (e.g., demographic data) are available. Future work could explore methods such as including auxiliary equations for $N(t)$ or assuming known functional forms to restore identifiability.\\
Another critical challenge in real-world applications is the uncertainty in initial conditions. Since initial values for $S(0)$, $E(0)$, $I(0)$, and $R(0)$ are often unknown or estimated from limited data, they can significantly impact parameter identifiability and estimation accuracy. Given these considerations, we consider analyses with and without knowing the initial conditions of the systems.\\

\begin{Proposition} \label{prop6}
The model~\eqref{eq:model6} is structurally identifiable from the observation of incidences, $y(t)= k E(t)$ with known initial conditions.
\end{Proposition}

\begin{table}[H]
	
    \caption{Summary of Structural Identifiability Results. This table summarizes the structural identifiability results for model parameters and the observability of state variables across six models: the generalized growth model (GGM), generalized logistic model (GLM), Richards model, generalized Richards model (GRM), Gompertz model (GOM), and SEIR model with inhomogeneous mixing. Results are presented for cases with and without known initial conditions. Parameters marked as "globally identifiable" can be uniquely determined from perfect data, while "locally identifiable" parameters require specific constraints for unique estimation. The observability of state variables, which indicates whether they can be inferred from the data, is also included. These findings were obtained using the StructuralIdentifiability.jl package in JULIA.}
\begin{tabular}{|c|c|c|c|c|}
\hline
\textbf{Model} & \textbf{Case}     & \textbf{Globally Identifiable}      & \textbf{Locally Identifiable}      & \textbf{Not Identifiable} \\ \hline
\textbf{GGM} \ref{ModelM1}& Without IC & $p, C(t), x(t)$ & - & -                     \\ \cline{2-5} 
& With IC & $p, C(t), x(t)$ & - & - \\ \hline

\textbf{GLM} \ref{ModelM2} & Without IC & $k, p, C(t), x(t)$&-& -\\ \cline{2-5} 
& With IC &  $k, p, C(t), x(t)$&-& - \\ \hline

\textbf{Richards} \ref{ModelM3}  & Without IC & $a, r, C(t), x(t)$&-& -\\ \cline{2-5} 
& With IC &  $a, r, C(t), x(t)$&-& - \\ \hline

\textbf{GRM} \ref{ModelM4}  & Without IC & $C(t)$&$a, p, x(t), z(t)$& -\\ \cline{2-5} 
& With IC &  $a, p, C(t), x(t), z(t)$&-& - \\ \hline

\textbf{GOM} \ref{ModelM5} & Without IC & $b, C(t), x(t)$&-& -\\ \cline{2-5} 
& With IC &  $b, C(t), x(t)$&-& - \\ \hline

\textbf{SEIR} \ref{ModelM6}  & Without IC & $k, \gamma, \alpha, S(t), E(t), I(t)$&-& $x(t),R(t),N$\\ \cline{2-5} 
& With IC &  $N, k, \gamma, \alpha, S(t), E(t), x(t), I(t), R(t)$&-& - \\ \hline
\end{tabular}
	 \label{SI_results}
\end{table}
\section{Fitting models to real epidemic data}

\subsection{Data}

We performed parameter estimation using the \textit{GrowthPredict} MATLAB Toolbox, a robust platform developed explicitly for fitting and forecasting time series trajectories based on phenomenological growth models using ordinary differential equations \cite{toolbox2024}. The models outlined in Equations~\eqref{eq:model2} to \eqref{eq:model6} were fitted to three sets of data: weekly incidence curve of Monkeypox data, COVID-19 data, and Ebola data described in the Data section. The computation of prediction intervals assumes that observed values are generated from the deterministic model with added normal error (mean zero, estimated variance). This assumption allows us to quantify uncertainty while maintaining consistency with the Monte Carlo framework used for practical identifiability analysis.

\subsection{Parameter estimation method}
We use the least squares (LSQ) method to perform the parameter estimation problem by minimizing the following objective function.

\begin{equation}\begin{split}
\bf\hat{p}_i=\mathop{\mathrm{argmin}}_{\bf{p_i}} \left(\sum_{j=1}^n (f(t_j, \bf{p_i})-y_{t_j})^2\right)^{1/2}.
\label{eq:objective_function}
\end{split}\end{equation}
Here, $f(t_j,\bf{p_i})$ is the predicted epidemic trajectory given by Model $i$, where $i=\{2,3,4,5,6\}$ and, $y_{t_j}$ is the time series data, where $t_j, j= 1,2,3,...,n$ are the discrete-time points for the time series data. This method was chosen for its computational efficiency and ability to handle non-linear parameter spaces effectively. 

To evaluate model performance, we use the \textit{GrowthPredict} MATLAB toolbox,which provides several goodness-of-fit metrics, including the Akaike Information Criterion (AIC), mean absolute error (MAE), mean squared error (MSE), and weighted interval score (WIS) \cite{toolbox2024}. The corresponding formulations are as follows.

Akaike information criterion (AIC) is computed as $$AIC = n \log(SSE)+2m+\D\frac{2m(m+1)}{n-m-1},$$
where $m$ is the number of model parameters, and $n$ is the number of data points (calibration period). The sum of squared errors $SSE$ is defined as
$$SSE= \sum_{j=1}^n (f(t_j, \bf{p_i})-y_{t_j})^2)^{1/2}.$$

Mean absolute error (MAE) and mean squared error (MSE) are given by
$$MAE=\frac{1}{n}\sum_{j=1}^n |f(t_j, \bf{p_i})-y_{t_j}|,$$
$$MSE = \frac{1}{n}\sum_{j=1}^n (f(t_j, \bf{p_i})-y_{t_j})^2.$$

The weighted interval score (WIS) is computed as
$$WIS_{\alpha_{0:K}}(F,y) = \D\frac{1}{K+\frac{1}{2}}(w_0 |y-\Tilde{y}|+ \sum_{k=1}^{K} w_k IS_{\alpha_k}(F,y)),$$
where $K$ is the number of prediction intervals (PIs), $\Tilde{y}$ is the predictive median, and $w_k=\frac{\alpha_k}{2}$ for $k=1,2,...,K$, with $w_0=1/2$. The interval scoring function is defined as, $$IS_{\alpha}(F,y)=(u-1)+\frac{2}{\alpha}(l-y)1(y<l)+\frac{2}{\alpha}(y-u)l(y>u),$$ where $1(y<l)$ is the indicator function that equals $1$ if $y<l$, and $0$ otherwise. The terms $l$ and $u$ represent the $\frac{\alpha}{2}$ and $1-\frac{\alpha}{2}$ quantiles of the forecast $F$, respectively.

The coverage of the $95\%$ prediction interval is given by,
$$95\% PI coverage = \frac{1}{N} \sum_{i=1}^{n}\mathbf{1}\{y_{t_i}>L_i\cap y_{t_i}<U_i\} \times 100 \%,$$
where $L_i$ and $U_i$ are the lower and upper bounds of the $95\%$ PIs, respectively, $y_{t_i}$ are the data and $\mathbf{1}$ is an indicator variable that equals 1 if $y_{t_i}$ is in the specified interval and 0 otherwise.

The estimated parameter values from fitting three sets of time series data; Monkeypox, COVID-19 and Ebola to models~\eqref{eq:model2}-\eqref{eq:model6} are given by Tables~\ref{tab:fitting_monkeyfox}, ~\ref{tab:fitting_covid} and~\ref{tab:fitting_ebola}, respectively.

\begin{table}[H]
\caption{Parameter estimates and model performance metrics obtained by fitting models~\eqref{eq:model2}-~\eqref{eq:model6} using the \textit{GrowthPredict} toolbox to weekly incidence curve of Monkeypox data.}

\vspace{1mm}
\begin{tabular}{c c c c  c c c c c c c c c c}
\hline
 Parameter&  $r$  & $\alpha$ & $k$ & $a$&$\beta$ & $\kappa$& $\gamma$ &AIC & MAE &MSE & WIS & Coverage  \\  
\hline

GLM&  $1.9$  & $0.84$   & $2.9e4$   & -& -&-&- &$432.76$ & $110.197$& $17245.91$& $63.16$ & 100\\ [1.5ex]   
\hline

RIC & $0.92$   & -  & $2.9e4$ & $0.36$&- &-&-&$405.67$ &$69.9$&$7691.72$&$43.16$ & 100\\ [1.5ex]   
\hline

GRM &   $2.3$ &$0.82$  &  $3.1e4$ & $0.9$ &-&-&- &$442.59$&$100.63$&$22044.17$&$69.72$& 93.75 \\ [1.5ex]   
\hline

GOM& $1.5$  &  -  &  -  & $0.2$ &-&-&-&$488.88$&$261.37$&$108193.6$&$160.41$ & 93.75\\ [1.5ex] 

\hline
SEIR&-&$0.96$&$4.7$&-&$7.3$&-&$4.8$&$407.19$&$60.8825$& $7296.921$&$40.749$&$96.875$\\[1.5ex]
\hline

\end{tabular}
\label{tab:fitting_monkeyfox} 
\end{table}

\begin{table}[H]
\caption{Parameter estimates and model performance metrics obtained by fitting models~\eqref{eq:model2}-~\eqref{eq:model6} using the \textit{GrowthPredict} toolbox to incidence curve of COVID-19 data.}

\vspace{1mm}
\begin{tabular}{c c c c c  c c c c c c c c c }
\hline
 Parameter&  $r$  & $\alpha$ & $k$ & $a$&$\beta$ & $\kappa$& $\gamma$ &AIC & MAE &MSE & WIS & Coverage  \\  
\hline

GLM&  $3.3$  & $0.7$   & $2e5$   & - &&&&$1755.59$& $512.55$&$387968.7$&$303.66$& 97\\ [1.5ex]   
\hline

RIC & $5$   & -  & $1.9e5$ & $0.089$& &&& $1761.20$&$485.79$&$410155.9$& $302.44$& 93\\ [1.5ex]   
\hline

GRM &   $4.8$ &$0.67$  &  $2.1e5$ & $1.2$ &&&&$1646.33
$&$764.37
$&$885696.74
$&$455.35$&$96.67$\\ [1.5ex]   
\hline

GOM & $0.93$  &  -  &  -  & $0.077$ &&&  &$1753.01$&$474.94$&$385462.9$&$294.35$& 93\\ [1.5ex]   
\hline
SEIR&-&$0.8$&-&-&$3.6$&$0.22$&$0.01$&$1910.94$&$1098.83$&$1903801$&$615.758$&$98$\\[1.5ex]
\hline
\end{tabular}
\label{tab:fitting_covid} 
\end{table}

\begin{table}[H]
\caption{Parameter estimates and model performance metrics obtained by fitting models~\eqref{eq:model2}-~\eqref{eq:model6} using \textit{GrowthPredict} toolbox to incidence curve of Ebola data.}

\vspace{1mm}
\begin{tabular}{c c c c c c c c c  c c c c c }
\hline
 Parameter&  $r$  & $\alpha$ & $k$ & $a$&$\beta$ & $\kappa$& $\gamma$ &AIC & MAE &MSE & WIS & Coverage  \\  
\hline

GLM&  $0.78$  & $0.85$   & $1.1e4$   & - & -&-&-&$783.6$& $3.472$&$1932.064$&$20.46$& 96.97 \\ [1.5ex]   
\hline

RIC & $0.46$   & -  & $1.1e4$ & $0.38$&- &-&-& $796.3$&$35.6$&$2318.422$& $22.25$&96.97 \\ [1.5ex]   
\hline

GRM &   $1.3$ &$0.96$  &  $1.3e4$ & $0.11$ &-&-&-&$857.57$&$60.2$&$5665.86$&$36.90$& 96.97\\ [1.5ex]   
\hline

GOM & $0.84$  &  -  &  -  & $0.1$ &-&-& &$860.2$&$63.97$&$6321.14$&$38.64$& 95.45\\ [1.5ex]   
\hline

SEIR&-&$0.99$&-&-&$5.7$&$5$&$5$&$816.345$&$43.813$&$3211.164$&$27.617$&$93.94$\\[1.5ex]

\hline
\end{tabular}
\label{tab:fitting_ebola} 
\end{table}

The figures below illustrate the fit of each model using the dataset that yielded the best fit based on AIC values.

\begin{figure}[H]
    \centering
    \begin{tabular}{c}
    \includegraphics[width=0.9 \linewidth]{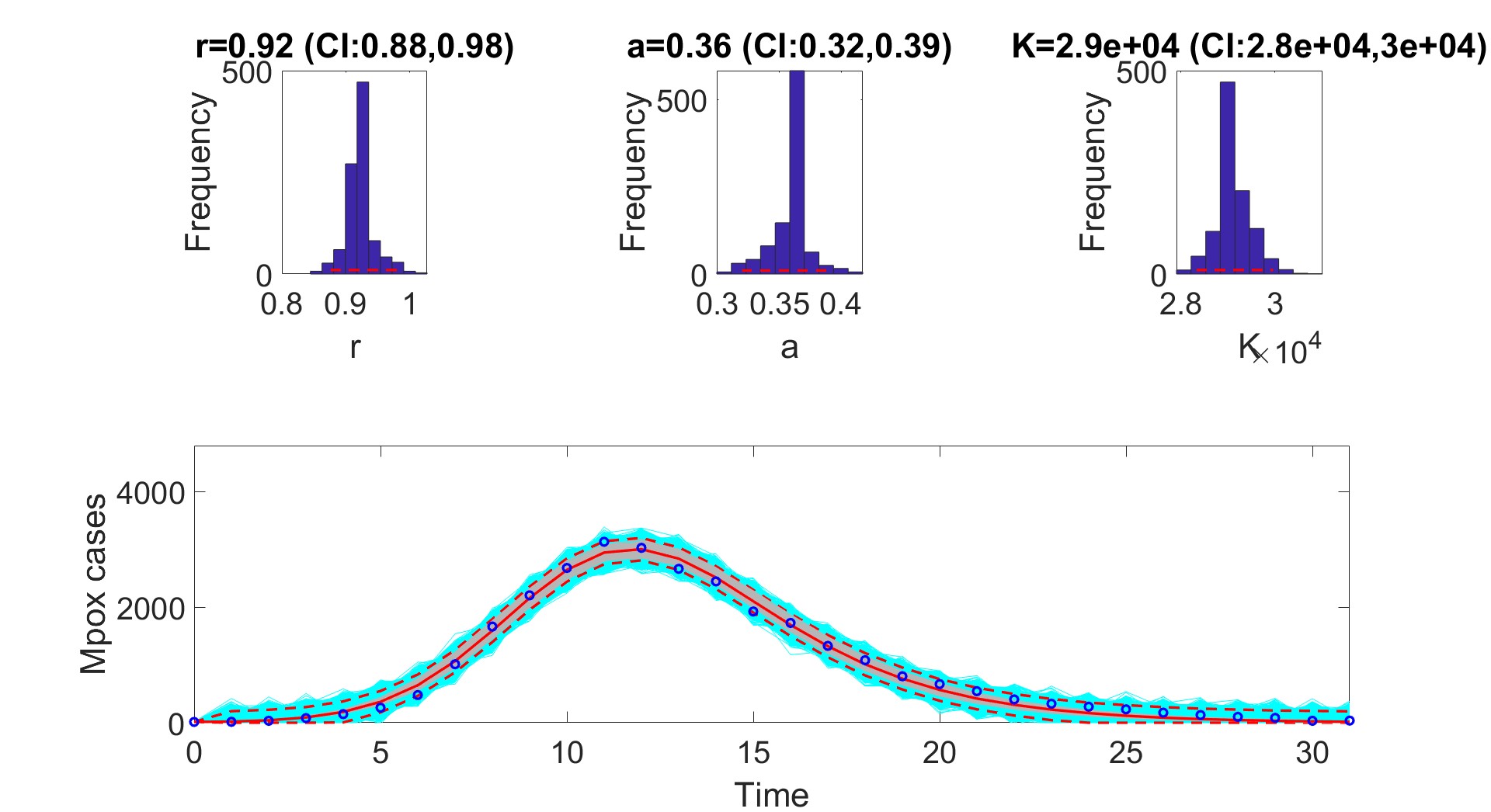} 
    \end{tabular}
    \caption{Fit of the Richards model to weekly incidence data for the Monkeypox epidemic. The model demonstrates a close alignment with observed data, capturing the non-linear growth dynamics characteristic of the outbreak. Key features such as the early exponential growth phase and subsequent plateau are well-represented, reflecting the model's robustness in describing sub-logistic epidemic patterns. The shaded region indicates the $95\%$ prediction intervals (PIs), highlighting the model's ability to quantify uncertainty in its projections.}
\end{figure}

\begin{figure}[H]
    \centering
    \begin{tabular}{c}
    \includegraphics[width=0.9 \linewidth]{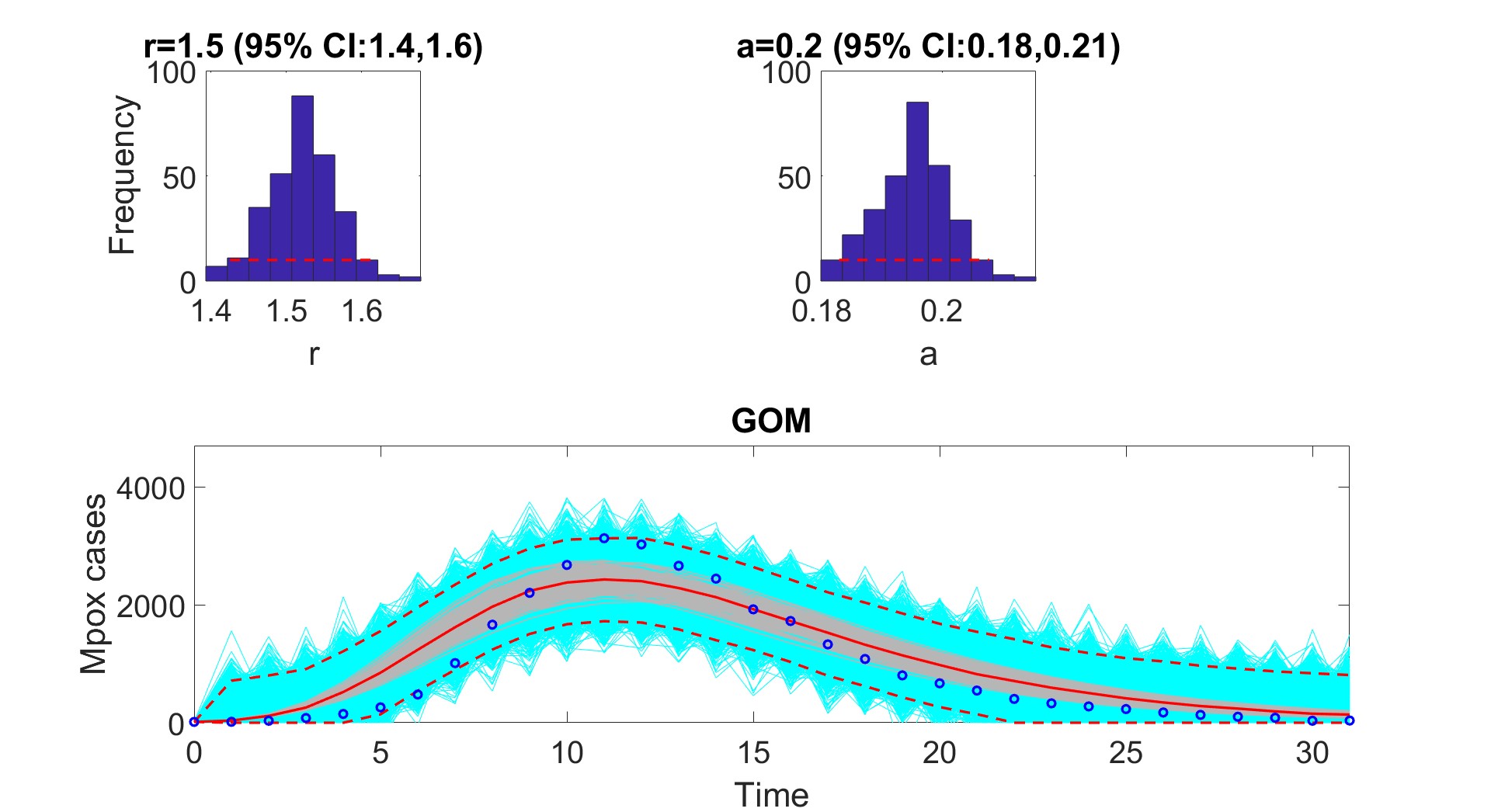} 
    \end{tabular}
    \caption{Model fit of the Gompertz model to the weekly incidence data for Monkeypox. The figure illustrates the observed weekly incidence data (dots) and the corresponding Gompertz model fit (solid line). The Gompertz model effectively captures the overall trend and deceleration in the growth of cases over time. Shaded regions represent the $95\%$ prediction intervals, highlighting the model's forecast uncertainty.}
\end{figure}

\begin{figure}[H]
    \centering
    \begin{tabular}{c}
    \includegraphics[width=0.9 \linewidth]{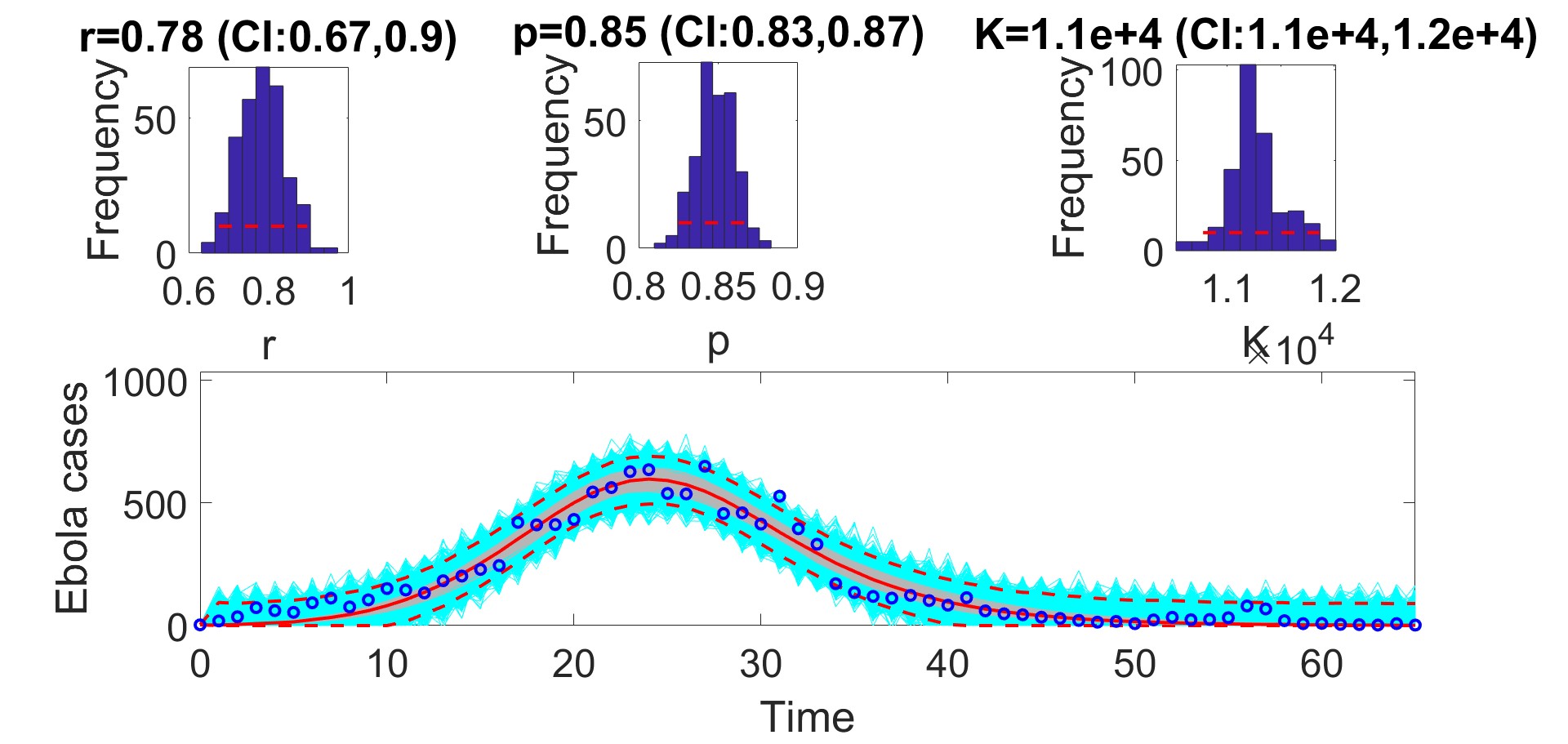} 
    \end{tabular}
    \caption{Model fit of the generalized logistic model (GLM) to the weekly incidence data for Ebola. The figure illustrates the alignment of the GLM with the observed data, showing its ability to capture the epidemic dynamics effectively. Key features of the GLM, such as its flexibility to model sub-exponential growth, are evident in the close agreement between the predicted trajectory and the observed data points. The model parameters were estimated using the least squares method, and the fit was evaluated based on the Akaike Information Criterion (AIC), mean absolute error (MAE), and mean squared error (MSE). The shaded region represents the $95\%$ prediction interval, highlighting the uncertainty in the model forecasts.}
\end{figure}

\begin{figure}[H]
    \centering
    \begin{tabular}{c}
    \includegraphics[width=0.9 \linewidth]{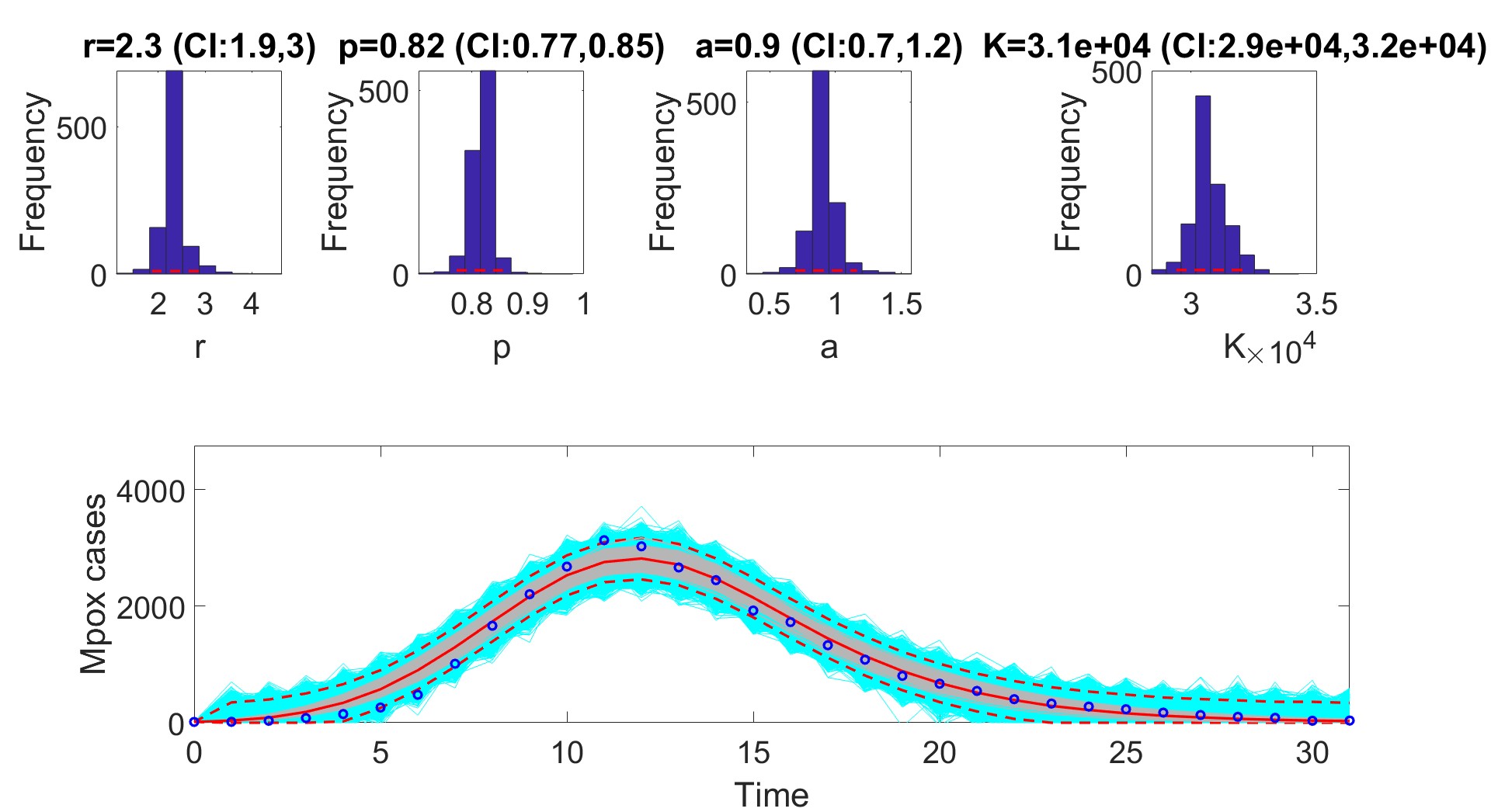} 
    \end{tabular}
    \caption{Fit of the generalized Richards model (GRM) to the weekly incidence curve of Monkeypox data. The figure illustrates the model's ability to capture the observed dynamics of the epidemic, highlighting the flexibility of the GRM in accommodating varying growth patterns. The best-fit parameter estimates were derived using the least squares method, and the model's performance is validated through metrics such as AIC, MAE, and MSE. Shaded areas represent the $95\%$ prediction intervals, demonstrating the model's uncertainty estimates and alignment with observed data points.}
\end{figure}

\begin{figure}[H]
    \centering
    \begin{tabular}{c}
    \includegraphics[width=0.9 \linewidth]{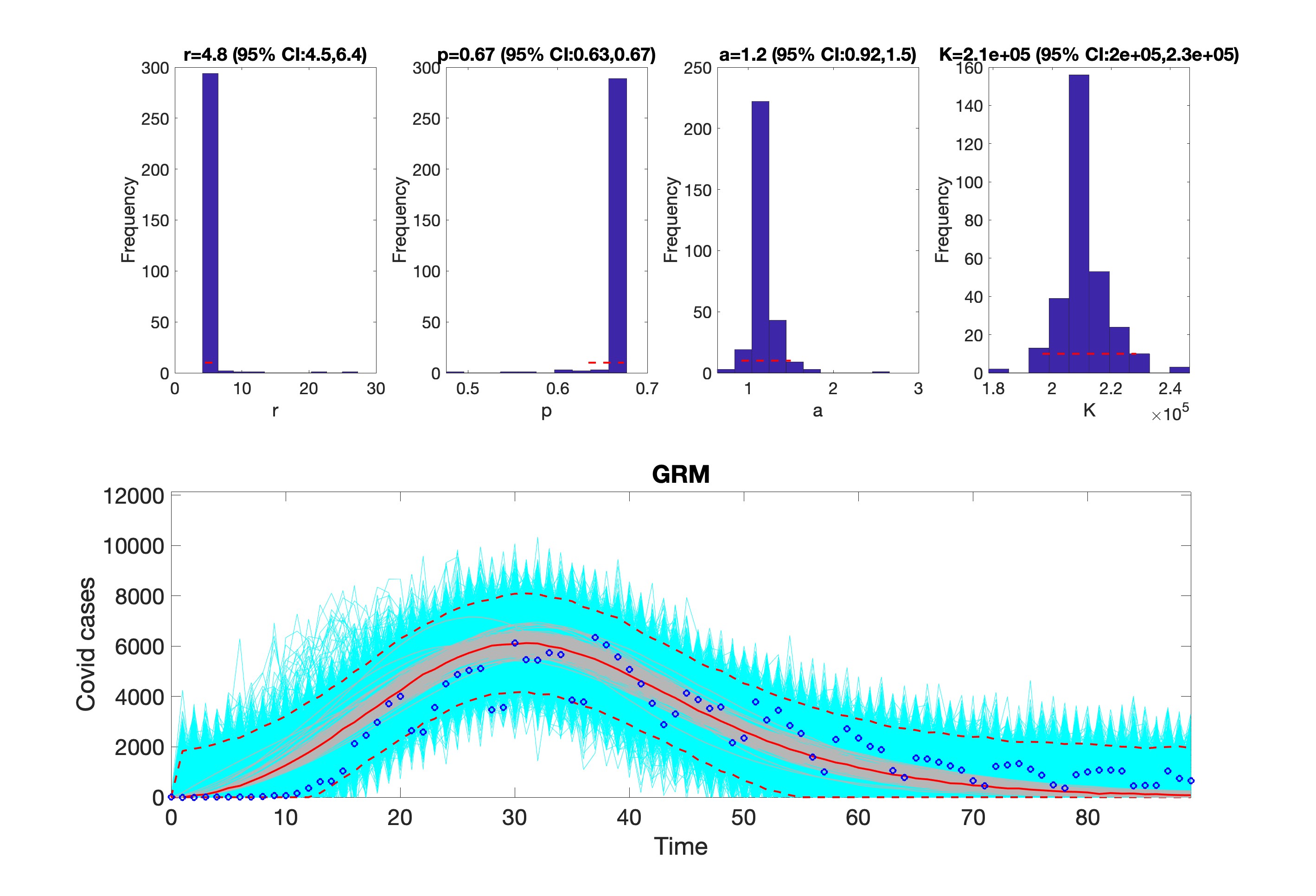} 
    \end{tabular}
    \caption{The fit of the generalized Richards model (GRM) to the COVID-19 dataset. This figure illustrates the model's ability to capture the dynamics of the epidemic, highlighting its flexibility in accounting for deviations from symmetric S-shaped growth curves. The GRM demonstrated the best fit for this dataset, as indicated by its low AIC score and high coverage of the $95\%$ prediction intervals. Observed data points are marked, while the model's predicted trajectory and uncertainty bounds are overlaid for comparison.}
\end{figure}

\begin{figure}[H]
    \centering
    \begin{tabular}{c}
    \includegraphics[width=0.9 \linewidth]{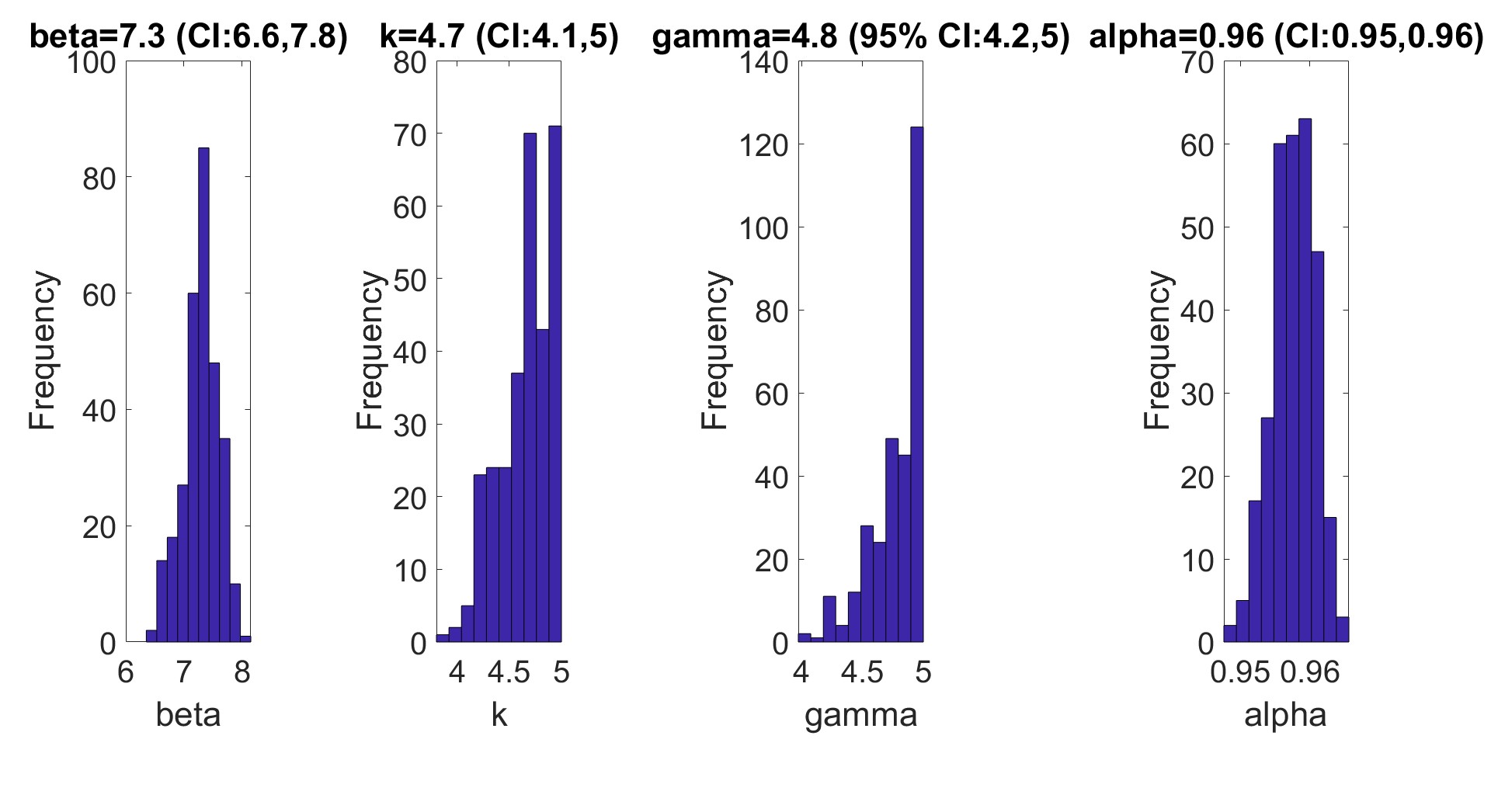} \\
    \includegraphics[width=0.9 \linewidth]{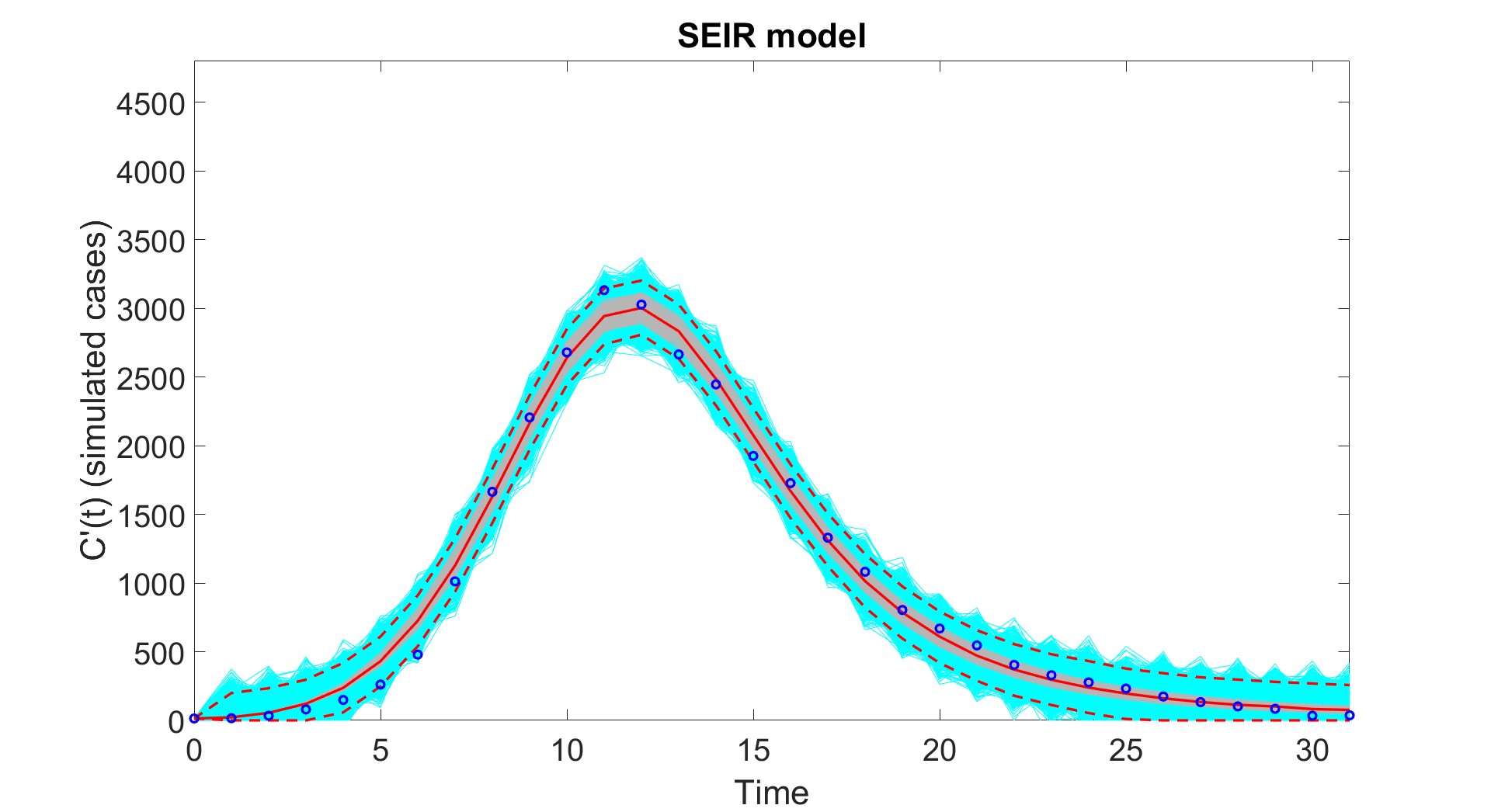} 
    \end{tabular}
    \caption{Fit of the SEIR model with inhomogeneous mixing to the Monkeypox dataset. The figure illustrates the model's ability to capture the weekly incidence trends, with parameter estimates derived from fitting the SEIR framework to observed data. Shaded regions indicate the $95\%$ prediction intervals, reflecting uncertainty in the model's forecasts. The model strongly aligns with the data, emphasizing its suitability for analyzing diseases with similar transmission dynamics. Key features of the fit, such as the peak incidence and decline phase, are accurately represented.}
\end{figure}

\section{Practical Identifiability}

Structural identifiability examines the feasibility of estimating parameters in a model with unlimited observation without measurement errors. However, real-world data are often collected discretely and may include significant measurement errors. Thus, the structural identifiability of a model does not imply the practically identifiable parameters. So, it becomes crucial to assess whether structurally identifiable parameters can be accurately estimated from data containing noise. Various methods are available to examine the practical identifiability \cite{miao2011,Miao2008,tuncer2018structural,Tuncer2018_zika,Wu2008,Wieland2021,Raue2014}. In this study, we employ the Monte Carlo simulation (MCS) method, described below \cite{miao2011,tuncer2018structural,Tuncer2018_zika,Miao2008}.

\begin{itemize}
    \item We solve the models Eq.~\eqref{eq:model2}- ~\eqref{eq:model6} numerically with the true parameter values $\bf{\hat{p}}$ obtained from the fitting of the model to experimental data from the data fitting section. Then we obtain the observations (output) at the experimental time points.
    \item We generate $M=1000$ data sets at each of the experimental time points with four different levels of increasing measurement error, $\sigma= 1\%, 5\%, 10\%, 20\%,$ by adding measurement errors to each given
experimental data point. Each measurement error is assumed to be distributed with a normal distribution with zero mean and standard deviation $\sigma$.
    \item We fit the model to each of the $1000$ data sets to estimate the parameter set $\bf{\hat{p}_i}$, using the objective function Eq. \eqref{eq:objective_function}.
    \item We calculate the average relative estimation error $(ARE)$ for each parameter in the model by,
 
\[ARE(\bf\hat{p}^{(k)}) = 100\% \times \frac{1}{M}\sum_{i=1}^M \frac{| \hat{p}^{(k)}-\hat{p}_i^{(k)} |}{|\hat {p}^{(k)}| } \]
where $\bf{\hat{p}}^{(k)}$ is the $k^{th}$ element of the true parameter set $\bf{\hat{p}}$ and $\bf{\hat{p}_i}^{(k)}$ is $k^{th}$ element of $\bf{\hat{p}_i}.$
   \item We use $AREs$ values to determine whether each of the model's parameters is practically identifiable, using definition~\ref{practical identifiability} given below.
   \end{itemize}

\begin{Definition}\label{practical identifiability}
Let $ARE$  be  the average relative estimation error of the parameter $\textbf{p}^{(k)}$. Let $\sigma$ be the measurement error.
\begin{itemize}
    \item if  $0< ARE(\hat{p}^{(k)})\le \sigma$ then parameter $\textbf{p}^{(k)}$ is strongly practically identifiable,
    \item if $\sigma< ARE(\hat{p}^{(k)})\le 10\sigma$ then parameter $\textbf{p}^{(k)}$ is weakly practically identifiable,
    \item if $ 10\sigma< ARE(\hat{p}^{(k)})$ then parameter $\textbf{p}^{(k)}$ is not practically identifiable.
\end{itemize}
A model is said to be practically identifiable if the parameters $\textbf{p}^{(k)}$ are practically identifiable, for all~$k$.
\end{Definition} 

In the data fitting section, we fitted the model equations Eq.~\eqref{eq:model2} through~\eqref{eq:model6} to three different datasets: Monkeypox data, Covid-19 data, and Ebola data. Here, we focus on assessing the practical identifiability of each model using the data set that produced the best fit based on lowest $AIC$ values. 
Thus, our investigation focuses on determining whether the \textbf{Richards} model, \textbf{GLM}, and \textbf{GRM} demonstrate practical identifiability when applied to the Monkeypox, Ebola, and COVID-19 datasets, respectively. Additionally, we evaluated the \textbf{GRM}, \textbf{GOM}, and \textbf{SEIR} models on the datasets where they demonstrated the best relative performance, to further assess their practical identifiability.

We found that all models, Richards, GRM, GOM, GLM, and SEIR, are practically identifiable from the given datasets. Each model's parameters were reliably estimated, with well-constrained confidence intervals, indicating that the data provided sufficient information to determine the parameters uniquely. Moreover, these results are consistent with the structural identifiability findings, further validating the models' capacity to capture the dynamics of the epidemics studied. The results of our analysis are presented in Table~\ref{tab:MCS_res_Rich_Mpox} for the Richard model, Tables~\ref{tab:MCS_res_GRM_Covid} and~\ref{tab:MCS_res_GRM_Mpox} for GRM, Table~\ref{tab:MCS_res_GLM_Ebola} for GLM, and Table~\ref{tab:MCS_res_GOM_Mpox} for GOM.

\begin{table}[H]
\begin{center}
\caption{Results of Monte Carlo simulations assessing the practical identifiability of the Richards model (Eq. 2.4) using virtual datasets generated at discrete Monkeypox experimental data points. The table presents the Average Relative Estimation Errors (AREs) for each model parameter ($r$, $a$, and $K$) under varying noise levels ($\sigma = 1\%$, $5\%$, $10\%$, $20\%$), alongside the corresponding confidence intervals (CI). These results demonstrate the model’s robustness in parameter estimation across different levels of observational noise.}

{\begin{tabular}{c c c c } \hline

 Parameter&  $r$  & $a$ & $K$   \\  \hline

$ARE$ \\$\sigma=1\%$ &  $2.8867\times 10^{-5}$  & $3.4759\times 10^{-5}$   & $4.1286\times 10^{-7}$   \\ [1.5ex] 

CI &$[0.9200, 0.9200]$ &$[0.3600, 0.3600]$& $[29000.0000, 29000.0000]$ \\
\hline

$ARE$ \\ $\sigma=5\%$ &  $8.5510\times 10^{-4} $  & $0.0015$   & $2.7180\times 10^{-4} $   \\ [1.5ex] 

CI & $[0.9200, 0.9200]$& $[0.3600, 0.3600]$& $[28999.6373, 29000.3469]$ \\
\hline 

$ARE$ \\ $\sigma=10\%$ &  $0.0020$  & $0.0036$   & $9.1805\times 10^{-4} $   \\ [1.5ex] 

CI & $[0.9200, 0.9200]$& $[0.3600, 0.3600]$& $[28999.2199, 29000.7637]$ \\

\hline  
$ARE$ \\ $\sigma=20\%$ &  $0.0042 $  & $0.0074 $   & $0.0021 $   \\ [1.5ex] 

CI & $[0.9199, 0.9201]$& $[0.3599, 0.3601]$& $[28998.3931, 29001.5577]$ \\

\hline  
\end{tabular}}
\label{tab:MCS_res_Rich_Mpox}

\end{center}
\end{table}

\begin{table}[H]
\begin{center}
\caption{Results of Monte Carlo simulations assessing the practical identifiability of the Generalized Richards Model (GRM, Eq.~\eqref{eq:model4}) using virtual datasets generated at discrete Monkeypox experimental data points. The table presents the Average Relative Estimation Errors ($ARE$s) for each model parameter ($r$, $\alpha$, $K$, and $a$) across varying noise levels ($\sigma = 1\%$, $5\%$, $10\%$, $20\%$). The corresponding confidence intervals (CI) are also provided, demonstrating the sensitivity of parameter estimates to different levels of observational noise.}

{\begin{tabular}{c c c c c} \hline

 Parameter&  $r$  & $\alpha$ & $K$ & $a$  \\  \hline

$ARE$ \\$\sigma=1\%$ &  $4.1031\times 10^{-6}$  & $9.0123 \times 10^{-7}$   & $1.2002\times 10^{-9}$ &$8.1173\times 10^{-6}$   \\ [1.5ex]  

CI & $[2.3000, 2.3000]$& $[0.8200, 0.8200]$& $[31000.0000, 31000.0000]$& $[0.9000, 0.9000]$ \\

\hline

$ARE$ \\ $\sigma=5\%$ &  $8.8096\times 10^{-4}$  & $2.0969\times 10^{-4}$   & $4.2781\times 10^{-5}$ & $0.0011$  \\ [1.5ex] 

CI & $[2.3000, 2.3002]$& $[0.8200, 0.8200]$& $[30999.9583, 31000.0272]$& $[0.9000, 0.9001]$ \\

\hline

$ARE$ \\ $\sigma=10\%$ &  $0.0039$  & $8.4508\times 10^{-4}$   & $3.3973\times 10^{-4}$ &$0.0043$  \\ [1.5ex] 

CI & $[2.2996, 2.3007]$& $[0.8200, 0.8200]$& $[30999.2908, 31000.7419]$& $[0.8999, 0.9002]$ \\

\hline

$ARE$ \\ $\sigma=20\%$ &  $0.0153$  & $0.0032$   & $0.0014$ & $0.0168$   \\ [1.5ex] 

CI & $[2.2984, 2.3018]$& $[0.8199, 0.8201]$& $[30997.8189, 31001.7471]$& $[0.8993, 0.9008]$ \\
\hline  
\end{tabular}}

\label{tab:MCS_res_GRM_Mpox} 
\end{center}
\end{table}

\begin{table}[H]
\begin{center}
\caption{Results of Monte Carlo simulations evaluating the practical identifiability of the Generalized Logistic Model (GLM, Eq.~\eqref{eq:model2}) using virtual datasets generated at discrete Ebola experimental data points. The table presents the Average Relative Estimation Errors ($ARE$s) for each model parameter ($r$, $a$, and $K$) under varying noise levels ($\sigma = 1\%$, $5\%$, $10\%$, $20\%$). These results quantify the model’s sensitivity to observational noise and highlight the robustness of parameter estimation.}

{\begin{tabular}{c c c c } \hline

 Parameter&  $r$  & $a$ & $K$   \\  \hline

$ARE$ \\$\sigma=1\%$ &  $4.5294\times 10^{-4}$  & $8.6649\times 10^{-5}$   & $5.4069\times 10^{-5} $   \\ [1.5ex]  

CI & $[0.7800, 0.7800]$& $[0.8500, 0.8500]$& $[10999.9599, 11000.0587]$\\

\hline

$ARE$ \\ $\sigma=5\%$ &  $0.0058 $  & $0.0011$   & $0.0019 $   \\ [1.5ex] 

CI & $[0.7799, 0.7801]$& $[0.8500, 0.8500]$& $[10999.4004, 11000.6236]$\\

\hline

$ARE$ \\ $\sigma=10\%$ &  $0.0126 $  & $0.0024 $   & $0.0043 $   \\ [1.5ex] 

CI & $[0.7798, 0.7803]$& $[0.8499, 0.8501]$& $[10998.7789, 11001.3100]$\\

\hline 

$ARE$ \\ $\sigma=20\%$ &  $0.0249 $  & $0.0047 $   & $0.0082 $   \\ [1.5ex] 

CI & $[0.7795, 0.7805]$& $[0.8499, 0.8501]$& $[10997.4842, 11002.4169]$\\

\hline  
\end{tabular}}

\label{tab:MCS_res_GLM_Ebola} 
\end{center}
\end{table}

\begin{table}[H]
\begin{center}
\caption{Monte Carlo simulation results for the Gompertz (GOM) model (Eq.~\eqref{eq:model5}) based on virtual datasets generated at discrete Monkeypox experimental data points. The table presents the Average Relative Estimation Errors ($ARE$s) for each model parameter across varying levels of observational noise ($\sigma = 1\%$, $5\%$, $10\%$, $20\%$). These results highlight the sensitivity of parameter estimates to noise and provide confidence intervals (CI) for each parameter to evaluate the robustness of the model under real-world data conditions.}
{\begin{tabular}{c c c } \hline

 Parameter&  $r$  & $a$  \\  \hline

$ARE$ \\$\sigma=1\%$ &  $0.0531 $  & $0.0533 $   \\ [1.5ex]  
CI & $[1.4980, 1.5019]$& $[0.1997, 0.2003]$\\

\hline 

$ARE$ \\ $\sigma=5\%$ &  $0.2639 $  & $0.2742 $     \\ [1.5ex] 

CI & $[1.4896, 1.5105]$& $[0.1987, 0.2014]$\\
\hline

$ARE$ \\ $\sigma=10\%$ &  $0.5309 $  & $0.5485 $   \\ [1.5ex] 

CI & $[1.4803, 1.5187]$& $[0.1974, 0.2027]$\\
\hline  
$ARE$ \\ $\sigma=20\%$ &  $1.0741 $  & $1.1507 $    \\ [1.5ex] 

CI & $[1.4596, 1.5384]$& $[0.1939, 0.2053]$\\
\hline  
\end{tabular}}

\label{tab:MCS_res_GOM_Mpox} 
\end{center}
\end{table}

\begin{table}[H]
\begin{center}
\caption{Monte Carlo simulation results for the Generalized Richards Model (GRM, Eq.~\eqref{eq:model4}) using virtual datasets generated at discrete COVID-19 experimental data points. The table presents the Average Relative Estimation Errors ($ARE$s) for each model parameter ($r$, $\alpha$, $K$, and $a$) under varying noise levels ($\sigma = 1\%$, $5\%$, $10\%$, $20\%$). Confidence intervals (CI) for the estimated parameters are also included, illustrating the impact of observational noise on parameter identifiability and estimation accuracy.}

{\begin{tabular}{c c c c c} \hline

 Parameter&  $r$  & $\alpha$ & $K$ & $a$  \\  \hline

$ARE$ \\$\sigma=1\%$ &  $0$  & $0$   & $0$ &$0$   \\ [1.5ex]  

CI & $[4.8,4.8]$& $[0.67, 0.67]$& $[21000.00, 21000.00]$& $[1.20, 1.20]$\\
\hline 

$ARE$ \\ $\sigma=5\%$ &  $4.9570 \times 10^{-9}$  & $1.5778\times 10^{-9}$   & $0$ &$4.0848\times 10^{-8}$  \\ [1.5ex] 

CI & $[4.8,4.8]$& $[0.67, 0.67]$& $[21000.00, 21000.00]$& $[1.20, 1.20]$\\
\hline

$ARE$ \\ $\sigma=10\%$ &  $2.8156\times 10^{-8}$  & $2.2234\times 10^{-9}$   & $0$ & $8.3800 \times 10^{-8}$  \\ [1.5ex] 

CI & $[4.8,4.8]$& $[0.67, 0.67]$& $[21000.00, 21000.00]$& $[1.20, 1.20]$\\
\hline  

$ARE$ \\ $\sigma=20\%$ &  $7.6621\times 10^{-8}$  & $3.5967\times 10^{-9}$   & $1.3859\times 10^{-17}$ & $4.0848\times 10^{-8}$   \\ [1.5ex] 

CI & $[4.8,4.8]$& $[0.67, 0.67]$& $[21000.00, 21000.00]$& $[1.20, 1.20]$\\
  
\hline  
\end{tabular}}

\label{tab:MCS_res_GRM_Covid} 
\end{center}
\end{table}

\begin{table}[H]

\caption{Results of Monte Carlo simulations assessing the practical identifiability of the SEIR model with nonhomogenous mixing (Eq. \eqref{eq:model6}) using virtual datasets generated at discrete Monkeypox experimental data points. The table presents the Average Relative Estimation Errors (AREs) for each model parameter ($\beta$, $\kappa$, $\gamma$, $N$, and $\alpha$) under varying noise levels ($\sigma = 1\%$, $5\%$, $10\%$, $20\%$). These results highlight the robustness of parameter estimation.}
{\begin{tabular}{c c c c c c} \hline

 Parameter&  $\beta$  & $\kappa$ &$\gamma$   & $N$ &$\alpha$ \\  \hline

$ARE$ \\$\sigma=1\%$ &  $5.3855\times 10^{-7}$  & $6.3303\times 10^{-8}$   & $3.0005\times 10^{-8}$ &$2.3138\times 10^{-15}$ &$1.0316\times 10^{-7}$  \\ [1.5ex]

CI & $[7.3000, 7.3000]$& $[4.7000, 4.7000]$&  $[4.8000, 4.8000]$ &$[100000.0000, 100000.0000]$& $[0.9600, 0.9600]$\\
\hline

$ARE$ \\ $\sigma=5\%$ &  $3.0360\times 10^{-4}$  & $3.0366\times 10^{-4}$   & $3.0385\times 10^{-4}$ & $3.3689\times 10^{-9}$ &$6.3055\times 10^{-5}$ \\ [1.5ex]

CI & $[7.3000, 7.3000]$& $[4.7000, 4.7000]$&  $[4.8000, 4.8000]$ &$[100000.0000, 100000.0000]$& $[0.9600, 0.9600]$\\
\hline

$ARE$ \\ $\sigma=10\%$ &  $0.0042$  & $0.0043$   & $0.0044$ &$4.1763\times 10^{-8}$ &$4.9886\times 10^{-5}$ \\ [1.5ex]

CI & $[7.2940, 7.3000]$& $[4.6999, 4.7041]$&  $[4.7960, 4.8000]$ &$[100000.0000, 100000.0000]$& $[0.9600, 0.9600]$\\
\hline

$ARE$ \\ $\sigma=20\%$ &  $0.0363$  & $0.0371$   & $0.0379$ & $9.5587\times e10^{-7}$&$2.6628\times 10^{-4}$   \\ [1.5ex] 

CI & $[7.2869, 7.3001]$& $[4.6997, 4.7089]$&  $[4.7909, 4.8000]$ &$[99999.9964, 100000.0022]$& $[0.9600, 0.9600]$\\
\hline  
\end{tabular}}

\label{tab:MCS_res_SEIR_Mpox} 
\end{table}

\section{Discussion}

This study evaluates the structural and practical identifiability of six commonly used phenomenological growth models in epidemiology. It demonstrates their robustness in parameter estimation and applicability across diverse datasets. To assess structural identifiability, we reformulated the models to address challenges posed by non-integer power exponents; we created an extended structure with fewer parameters and additional equations while preserving the original degrees of freedom. Our findings confirm that all reformulated models are structurally identifiable, even with unknown initial conditions. The original model formats were applied for data fitting and practical identifiability analyses, revealing that all models remained practically identifiable under varying noise levels, with performance differing across datasets.

The first step in validating these models involved assessing whether all unknown parameters were structurally identifiable, ensuring they could theoretically be uniquely determined from perfect, unlimited data. This step establishes a necessary foundation for reliable parameter estimation. We used the differential algebra method for this evaluation \cite{tuncer2018structural,Chowell2023,Tuncer2018_zika}. Given the lack of dedicated software tools for analyzing systems of ordinary differential equations with non-integer exponents, we reformulated the models by introducing additional state variables. This reformulation enabled the use of the StructuralIdentifiability.jl package in Julia \cite{structidjl2023}. Our findings demonstrated that all parameters in the reformulated models were structurally identifiable, even with unknown initial conditions. By assessing the observability of state variables, we inferred the structural identifiability of the original models, bridging the gap between the reformulated and original structures \cite{Margaria2004,Margaria0011}.

Model validation was conducted using the \textit{GrowthPredict} MATLAB Toolbox, a specialized tool for fitting and forecasting time-series trajectories based on phenomenological growth models. This toolbox facilitated comparisons across datasets and enabled the calculation of performance metrics such as AIC, MAE, MSE, and WIS. This toolbox was applied to three epidemiological datasets: weekly incidence data for monkeypox, COVID-19, and Ebola \cite{toolbox2024}. To compare performance, metrics such as AIC, MAE, MSE, WIS, and coverage were calculated for each model. The selection of models based on AIC values revealed that certain models are better suited to specific epidemic contexts. For example, the Richards model provided the best fit for Monkeypox data. In contrast, the GRM model excelled at fitting the COVID-19 data. The GLM model was most effective for the Ebola data.

In the next phase, we assessed practical identifiability by applying the models to datasets that produced the best fits, as indicated by AIC values. We focused on the Richards model, SEIR, and Gompertz models with the Monkeypox dataset, the GLM with the Ebola dataset, and the GRM with the COVID-19 dataset. Practical identifiability was evaluated using Monte Carlo simulations, highlighting the robustness of parameter estimation under real-world conditions, where data is often noisy and limited. Other approaches, such as the Fisher Information Matrix (FIM), Profile Likelihood Method, and Bayesian methods, could complement this analysis \cite{miao2011,tuncer2018structural,Tuncer2018_zika,Miao2008}. Our results showed that all the models were practically identifiable for their respective datasets. However, parameter estimation accuracy varied across models and datasets, emphasizing the sensitivity of these models to data quality. These findings underscore the importance of including practical identifiability analysis in epidemiological studies to ensure robust model validation.

By addressing the challenge of non-integer power exponents and validating the models under practical conditions, this study broadens the applicability of phenomenological models. It strengthens confidence in their use for epidemic forecasting. However, several limitations must be acknowledged. First, while introducing additional state variables facilitates structural identifiability analysis, further research is needed to assess its impact on computational efficiency and model interpretability. Second, the dependency of parameter estimation accuracy on data quality remains a challenge, emphasizing the need for high-quality, well-calibrated datasets. Third, alternative approaches such as Bayesian methods and Fisher Information Matrix analysis should be explored to complement the Monte Carlo simulations performed here. Finally, future research should examine the role of time-varying parameters and unknown initial conditions in shaping identifiability outcomes.

Our analysis highlights the structural and practical identifiability of six phenomenological models across various epidemiological datasets. The results underscore the utility of these models in forecasting disease dynamics and their adaptability to diverse epidemic contexts. However, the accuracy of parameter estimates is highly dependent on data quality, emphasizing the critical need for careful data collection and the integration of practical identifiability evaluations in the model selection process.

\vspace{6pt} 


\section*{Author Contributions}Conceptualization, Y.R.L, G.C, G.P. and N.T.; methodology, Y.R.L, G.C, G.P. and N.T.; software, Y.R.L, G.C, G.P. and N.T.; validation, Y.R.L, G.C, G.P. and N.T.; formal analysis, Y.R.L.; investigation, Y.R.L.; writing---original draft preparation, Y.R.L, G.C, G.P. and N.T.; writing---review and editing, Y.R.L, G.C, G.P. and N.T.; visualization, Y.R.L; supervision, Y.R.L, G.C, G.P. and N.T.; methodology, Y.R.L, G.C, G.P. and N.T. All authors have read and agreed to the published version of the manuscript.

\section*{Funding} G.C. is partially supported by NSF grants 2125246 and 2026797. G.P. is supported by the French ANR-22-CE48-0008 OCCAM project. N.T. and Y.R.L. are supported by NIH NIGMS grant no. 1R01GM152743-01.

\section*{Data Availability} The data for the research is available at \url{https://github.com/YuganthiLiyanage/Phenomenological-Growth-Models}

\section*{Conflicts of Interest} The authors declare no conflicts of interest.

\bibliographystyle{unsrt}
\bibliography{References}

\end{document}